\documentclass[superscriptaddress,twocolumn,showpacs,prl,amsmath,amssymb]{revtex4}
\usepackage[varg]{txfonts}

\usepackage{graphicx}
\usepackage{bm}

\def\XXint#1#2#3{{\setbox0=\hbox{$#1{#2#3}{\int}$}
    \vcenter{\hbox{$#2#3$}}\kern-.5\wd0}}

\begin{document}

\title{Spin-triplet pairing instability of the spinon Fermi surface
in a $U(1)$ spin liquid}

\author{Victor~Galitski}
\affiliation{Department of Physics and Joint Quantum Institute,
University of Maryland, College Park, MD 20742-4111}
\author{Yong~Baek~Kim}
\affiliation{Department of Physics, University of Toronto,
Toronto, Ontario M5S~1A7, Canada}

\begin{abstract}
Recent experiments
on the organic compound $\kappa$-$(ET)_2Cu_2(CN)_3$ have provided
a promising example of a two dimensional spin liquid state. This
phase is described by a two-dimensional spinon Fermi sea coupled
to a $U(1)$ gauge field. We study Kohn-Luttinger-like pairing
instabilities of the spinon Fermi surface due to singular
interaction processes with twice-the-Fermi-momentum transfer. We
find that under certain circumstances the pairing instability
occurs in odd-orbital-angular-momentum/spin-triplet channels.
Implications to experiments are discussed.
\end{abstract}

 \maketitle

Understanding possible phases of matter in strongly interacting
electron systems is one of the central issues in condensed matter
physics. A prime example of such systems is the Mott insulator.
Among a plethora of possible insulating phases, perhaps the most
interesting states are spin liquids with no long range order.
The possibility of such states of matter was first discussed by
Pomeranchuk back in 1941~\cite{Pomeranchuk}, who conjectured that
in insulators, the elementary excitations may be charge-neutral
fermions (spinons). More precise and modern form of such proposals
had to wait until Anderson proposed the spin
liquid state as the key paradigm in high-$T_c$ cuprates physics ~\cite{Anderson}.

While the relevance of spin liquid phases to the high-$T_c$
problem is still under debate, the search for spin liquids has
continued in other classes of strongly interacting electron
systems. One of the most notable examples is a series of recent
experiments on the organic compound
$\kappa$-$(ET)_2Cu_2(CN)_3$~\cite{Shimizu_etal,Kurosaki_etal}.
Here the electrons are half-filled on the triangular lattice and
the relative interaction strength is controlled by applying a
hydrostatic pressure. The insulating phase of this system exhibits
finite uniform susceptibility and finite specific heat coefficient
at low temperatures, {\it i.e.}~metal-like behaviors. It was
proposed that the $U(1)$ spin liquid phase with a spinon Fermi
surface is a viable explanation of the experimental results
\cite{Lesik,SSLee}. Variational calculations on the
Heisenberg-ring-exchange model are consistent with this
expectation~\cite{Lesik}.

The spinon Fermi sea, however, is strongly coupled to a $U(1)$
gauge field. On the lattice, the $U(1)$ gauge field is compact and
in principle one should worry about possible confinement effect in
two dimensions due to monopole events~\cite{Polyakov1}. It has
been argued that the monopole events are suppressed due to the
coupling to the gapless spinon degrees of freedom
\cite{IL,Mike_etal}. Nevertheless, the gauge field still gives
rise to singular renormalization of various physical quantities.
For example, the specific heat coefficient should diverge as
$T^{-1/3}$ \cite{Nayak_Wilczek,YBK_etal, AIM, Polchinski,mymetal}.
This behavior has not been seen and instead there exist abrupt
changes in the susceptibility as well as in the specific heat
around a few Kelvin \cite{Shimizu_etal,Kurosaki_etal}. This may
suggest a thermodynamic transition or possibly a crossover. It is
therefore interesting to consider possible instabilities of the
corresponding spinon Fermi sea state.

The instability of the spinon Fermi surface may also be a very
useful avenue to study the emergence of other possible phases.
Notice that in Fermi liquid theory, all broken symmetry states of
ordinary metals can be understood as an instability of the Fermi
surface. In the same spirit, the spinon Fermi surface state may be
regarded as a mother state of various possible phases of Mott
insulators.

In this paper, we study the Kohn-Luttinger-like pairing
instabilities~\cite{KL} of the spinon Fermi surface. Due to the
coupling to the gapless gauge field, the spinon interaction vertex
for the momentum transfer of twice the Fermi momentum diverges as
a power law \cite{AIM}. This singularity renormalizes the
effective interaction in the Cooper channel, which becomes {\it
attractive} for odd orbital angular momenta. It is found that
under certain conditions the system is unstable to the spinon
paired state in odd orbital angular momentum (spin-triplet)
channels.  When the pairing is present, the ground state would be
a $Z_2$ spin liquid.

Some remarks on the relation to the experimental findings are in
order. a) As mentioned earlier, the singular temperature
dependence expected from the gauge interaction has not been seen
at low temperatures. This is consistent with the fact that the
spinon pairing  gaps out the $U(1)$ gauge field below a pairing
temperature so that various singularities associated with the
$U(1)$ gauge field fluctuations do not show up at low
temperatures. b) The specific heat does not depend on the applied
field up to 8T \cite{Nakazawa_etal}. This is consistent with the
spin-triplet pairing that is not affected by the Zeeman effect. c)
When the system becomes superconducting at higher pressure, the
Knight shift does not change across the superconducting transition
\cite{Shimizu_etal2}. This can be explained if the resulting
superconducting state is related to the spinon pairing state and
hence a spin-triplet superconductor.

Previously, Lee et al.~\cite{lees} proposed a different mechanism
for spinon pairing. It is based on the ``Amperian pairing'' that
arises due to the attractive current-current interaction between
spinons moving in the same direction. Such an interaction leads to
Cooper pairs with a finite center of mass momentum, resulting in a
translational-symmetry broken state. The main difference between
our proposal and that of Ref.~[\onlinecite{SSLee}] is that our
spin-triplet state occurs in the normal BCS channel and
corresponds to a uniform ground state.

We start with the following Hamiltonian, which describes a system
of fermionic spinons minimally coupled to a $U(1)$ gauge
field~\cite{IL,Nayak_Wilczek,YBK_etal,AIM,Polchinski,mymetal}:
\begin{eqnarray}
\label{H1} {\cal H}_{a,f} = \sum_{s} \int d^2r \ f^{\dagger}_{s}
({\bf r}) \ \epsilon (-i{\bm \nabla} - {\bf a}) \ f_{s} ({\bf r})
\end{eqnarray}
where $\epsilon (-i{\bm \nabla} - {\bf a})$ is obtained by
replacing ${\bf p}$ by $(-i{\bm \nabla} - {\bf a})$ in the spinon
dispersion $\epsilon ({\bf p})$. Here $f^\dagger$ and $f$ are the
spinon creation and annihilation operators. $s = 1,2,\ldots N$ is
the ``spin'' index generalized to $N$ components, and ${\bf a}$ is
the fluctuating gauge field.  The bare gauge field Hamiltonian can
be taken as ${\cal H}_a = \frac{1}{2} \int d^2 {\bf r} \left\{
g_0^2 {\bf e}^2({\bf r}) + \frac{1}{g_0^2} \left[ {\bm \nabla}
\times {\bf a}({\bf r}) \right]^2 \right\},$ where ${\bf a}$ and
${\bf e} = - i \frac{\partial}{\partial {\bf a}}$ are canonically
conjugate quantum operators, and $g_0$ is a bare gauge coupling
constant. However, the exact form and parameters of the bare gauge
field theory are unimportant, as the coupling to spinons generates
more relevant terms. This is pictorially described in Fig.~1.1a,
where the dynamics of the field is determined by the particle-hole
excitations of spinons via the RPA renormalization. The resulting
effective propagator of the transverse gauge field takes the
following form
\begin{equation}
\label{clean_propagator} {\cal D}_{\alpha\beta}(\omega,{\bf q}) =
P_{\alpha\beta}^{({\rm tr})} ({\bf q}) \frac{1} {-i\gamma
\omega/q + \chi q^2},
\end{equation}
where $P_{\alpha\beta}^{({\rm tr})} ({\bf q}) =
\delta_{\alpha\beta} - q_\alpha q_\beta/q^2$, and $\gamma$ and
$\chi$ are constants, which determine the Landau damping and the
Landau diamagnetic susceptibility of the fermions, respectively.
In the large-$N$ limit, one can develop a self-consistent Eliashberg-type
\cite{Eliashberg} theory by using the electronic
self-energy (Fig.~1.1b), $\Sigma (\varepsilon) = i\varepsilon
\left| \omega_0/\varepsilon \right|^{1/3}$ (where $\omega_0 \sim
E_F$ is a constant). This leads to a non-Fermi liquid behavior.
The self-consistency of the theory implies that (i)~The lines in
the bosonic and fermionic self-energies in Figs.~1.1a and 1.1b can
be either thin (bare propagator) or thick (renormalized propagator),
leading to the same result.
Further renormalization does not change the effective dynamics of
the fermions and gauge bosons. (ii)~Once the above renormalization
is taken into account, the vertex corrections are
small (see Fig.~1.2). The latter statement is indeed true in the
large-$N$ limit, but only if the momentum transfer is not close to
$2 p_F$. As pointed out by Altshuler et al.~\cite{AIM}, the
vertex diverges logarithmically if the momentum transfer is
exactly equal to twice the Fermi momentum.
The leading logarithms can be summed within the standard parquet
technique leading to a power-law divergence of the vertex
\begin{equation}
\label{G2p} \Gamma_{2 p_F} (\omega,{\bf q})= {\Gamma_0 \over
\left[ \left| {\omega / E_F} \right| + c \left( {q/2 p_F} - 1
\right)^{3/2} \right]^\sigma},
\end{equation}
where $\Gamma_0$ is the value of the vertex far from the
$2p_F$-anomaly and $c$ and $\sigma$ are some $N$-dependent
constants. In Ref.~\cite{AIM}, it was shown that within the
large-$N$ treatment, $\sigma \propto 1/N$. The extrapolation of
the results in the large-$N$ limit to $N=2$ leads to $\sigma
\approx 0.36$ for the circular Fermi surface. It can be shown that
$\sigma$ increases as the curvature of the Fermi surface becomes
larger \cite{KM}. On the other hand, the small-$N$ approach in
Ref.~\cite{AIM} gives $\sigma \approx 0.52$. Clearly, these
results provide only an estimate of the exponent $\sigma$ as there
is no truly controlled method of treating strong gauge
fluctuations.

\begin{figure}[htbp]
\centering
\includegraphics[width=3.1in]{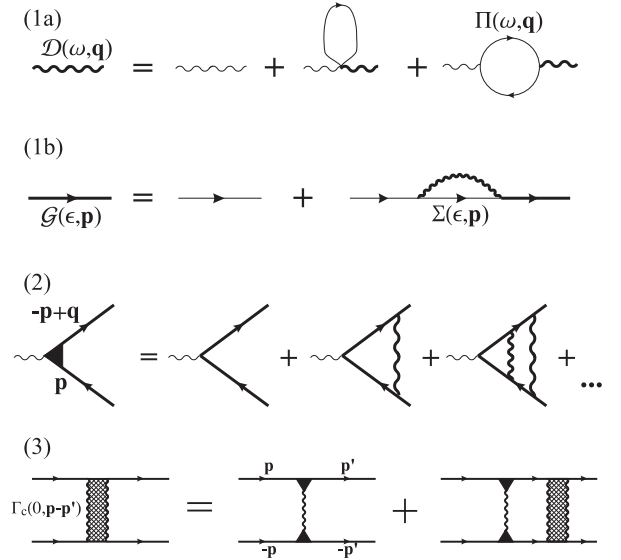}
\caption{(1a): The RPA approximation for the gauge-field
propagator. (1b): The leading contribution to the fermionic
self-energy. (2): The renormalization of the spinon vertex by
gauge interactions. The corresponding diagrams diverge
logarithmically if the momentum transfer is equal to two Fermi
momenta, $q=2p_F$. (3): The Bethe-Salpeter equation for the
irreducible part of the scattering amplitude in the Cooper
channel. The wavy line includes all interactions, including the
magnetic interactions mediated by the gauge field. The shaded
vertices represent processes described by Fig.~1.2.}
\label{fig1}
\end{figure}

Here we assume that
the structure of the theory found at large
$N$ is preserved in the physical limit of $N = 2$.
This implies that the structures of all power laws in the
bosonic and fermionic propagators are the same but
the parameters of the effective theory can be
some non-universal numbers.
One of these parameters is
the exponent $\sigma$ in Eq.~(\ref{G2p}). This parameter
also enters the temperature dependence of the susceptibility and
therefore is a physically observable quantity.
Thus instead of describing the
effective theory in terms of the unphysical number of flavors $N$,
it is perhaps more reasonable to use the parameter $\sigma$ as the
variable which ``controls'' the effective theory. This is the
point of view we take in this paper.

We note that the power law divergence (\ref{G2p}) can be viewed as
a strong renormalization of the Kohn anomaly in the non-Fermi
liquid phase. A natural question is whether there are
instabilities related to this anomaly. Ref.~\cite{AILM}
studied possible instabilities in the particle-hole channel and
concluded that a strong enough short-range interaction may lead to
a density-wave transition in the system. Below we address the
issue of possible instabilities in the Cooper channel. The
theoretical motivation comes from the following observation: In  a
usual Fermi liquid, the effective interaction with momentum $2
p_F$, is not divergent, but non-analytic. This weak
non-analyticity is not benign and leads to a number of
observable effects such as ``non-Fermi-liquid'' temperature
corrections to thermodynamics and
transport coefficients~\cite{VG_SDS,VG_Chubukov,Chubukov_etal}.
This anomaly can also give rise to
superconducting or pairing instabilities even in
the presence of repulsive interactions.
The latter phenomenon has been known as the
Kohn-Luttinger effect~\cite{KL,Chubukov,VG_SDS_KL}.
In a usual Fermi liquid,
this effect arises because the effective interaction
in real space acquires an oscillatory part (Friedel oscillations),
which is a reflection of the Kohn anomaly. Thus there appear
attractive regions and electrons can pair up due to this
attraction. They however, must be far apart from each other which
implies a finite orbital angular momentum of a pair. In a gauge-fermion
system, the $2 p_F$ anomaly is much stronger than in the ordinary
Fermi liquid, thus it is natural to consider the possibility of
spinon pairing of Kohn-Luttinger-type.

Fig.~1.3 shows pictorially the Bethe-Salpeter equation for the
effective interaction in the Cooper channel. Below we will
concentrate on the case when the momentum transfer is of order $2
p_F$. The wavy line in Fig.~1.3 is assumed to be a combination of
the effective magnetic interaction and short-range repulsive
interactions which may include the screened ``electric'' forces
mediated by the gauge field. From now on, we will denote the
corresponding interaction as $U(2 p_F)$. It is important to
include the singular $2p_F$ vertices in the Bethe-Salpeter
equation. The resulting equation has the following form:
\begin{eqnarray}
\label{BS}  \mathcal{T}_c \left(\varepsilon,\varepsilon'; {\bf
p},{\bf p}'\right) = \nonumber  &&\!\!\!\!\! U (\varepsilon -
\varepsilon',{\bf p} - {\bf p}') \Gamma_{2 p_F}^2(\varepsilon -
\varepsilon',{\bf p} - {\bf p}')  \\ && \!\!\!\!\!\!\!\!\!\!
\!\!\!\!\!\!\!\!\!\! - T \sum_{\eta} \int_{\bf k} U ( \varepsilon
-\eta,{\bf p} - {\bf k} ) \Gamma_{2 p_F}^2 ( \varepsilon
-\eta,{\bf p} - {\bf k} ) \nonumber\\ \!\!\!\!\! &&\times\,\,
{\cal G}_{\eta} ({\bf k}) {\cal G}_{-\eta} (-{\bf k})
\mathcal{T}_c \left(\eta,\varepsilon'; {\bf k},{\bf p}' \right),
\end{eqnarray}
As usual in the Cooper problem, we assume that the integral on the
right-hand side is determined by momenta at the Fermi surface and
small frequencies $\eta \to 0$. Therefore, the momentum dependence
of all functions in Eq.~(\ref{BS}) reduces to the dependence on a
single angular variable (i.e., the angle between two of the
following vectors ${\bf p}$, ${\bf p}'$, and ${\bf k}$). We
therefore can simplify the Bethe-Sapleter equation by performing a
standard decomposition into harmonics corresponding to different
orbital angular momenta, i.e. $f_l = \int_0^\pi f(\phi) \cos{(l \phi)}
d\phi/\pi$, where $f(\phi)$ is an arbitrary function of the angle.
In these notations, we can write Eq.~(\ref{BS}) as
\begin{eqnarray}
\label{BS2}  \mathcal{T}_c \left(\varepsilon,\varepsilon'; l
\right) = \nonumber  &&\!\!\!\!\! U (2 p_F) \left[ \Gamma_{2
p_F}^2 \right] (\varepsilon - \varepsilon',l)  \\ &&
\!\!\!\!\!\!\!\!\!\! \!\!\!\!\!\!\!\!\!\! - U (2 p_F) T
\sum_{\eta}  \left[ \Gamma_{2 p_F}^2 \right] ( \varepsilon -
\eta,l) C(\eta) \mathcal{T}_c \left(\eta,\varepsilon'; l \right),
\end{eqnarray}
Here $C(\eta)$ is the Cooperon and we assumed that the interaction
(without the vertex corrections) has no singularity at $q =
2 p_F$.
Pairing instability arises in principle only if there is an
effective attraction in one of the channels labelled by the
orbital angular momentum $l$. In the large-$N$ limit, the
parameter $\sigma$ may be small and by performing the Fourier
transform of the square of the $2 p_F$-vertex, we obtain ($\sigma
< 1/6$)
\begin{equation}
\label{s<1/6} \left[ \Gamma_{2 p_F}^2 \right] \left(0,l\right) =
{2^{6\sigma} \Gamma_{0}^2 \over (1-6\sigma)} B^{-1} (1-3\sigma
+l,1-3\sigma-l) \propto {(-1)^{l+1} \over l^{1-6\sigma}},
\end{equation}
where $B(p,q) = \Gamma(p+q)/\left[\Gamma(p) \Gamma(q)\right]$ is
the beta-function and the last estimate on the right-hand side
corresponds to the limit $l \gg 1$. Therefore, for small $\sigma <
1/6$, the effective interaction is attractive for large even
orbital angular momenta. However, in the physical limit of $N = 2$, there
is no reason for the parameter $\sigma$ to be small. If $\sigma >
1/6$, the corresponding Fourier transform of the static
double-vertex diverges, but it is cut off by the frequency and we
get ($\sigma > 1/6$)
\begin{equation}
\label{s>1/6} \left[ \Gamma_{2 p_F}^2 \right]
\left(\omega,l\right) = {2 \Gamma_{0}^2  B(1/3,2\sigma-1/3) \over
3 \pi c^{1/3}} \left( -1 \right)^l \left| {E_F \over \omega}
\right|^{2\sigma - {1 \over 3}},
\end{equation}
In this regime, the effective interaction is attractive for all
odd orbital angualr momenta, $l = 1, 3, 5, \ldots$.

We now address the possibility of a pairing instability due to the
effective attraction. In the usual Fermi liquid an attractive
interaction automatically implies a pairing instability, which is
a consequence of the Cooper logarithmic divergence. It is not so
in the fermion-gauge system because the Cooperon divergence is
much weaker here than in the Fermi liquid [we recall that ${\cal
G}_{\eta} ({\bf k}) = ( i \eta |\omega_0/\eta|^{1/3} - \xi_{\bf
p})^{-1}$]: $ C(\eta) = \int {d^2 {\bf k} \over (2 \pi)^2} {\cal
G}_{\eta} ({\bf k}) {\cal G}_{-\eta} (-{\bf k}) = {\nu \pi /
(|\eta|^{2/3} \omega_0^{1/3})}$, where $\nu$ is the density of
states.

To find an instability, we need to consider the following
 BCS self-consistency equation
\begin{equation}
\label{BCSeq} T U(2 p_F)\sum_{\eta} \left[ \Gamma_{2 p_F}^2
\right] \left(\varepsilon - \eta,l\right) C(\eta) \Phi_l(\eta) =
 \Phi_l(\varepsilon),
\end{equation}
where $l$ is the orbital angular momentum and $\Phi_l$ is related
to the corresponding pairing amplitude. The existence of a
non-trivial solution of the corresponding eigenvalue problem
implies  a divergence of the resolvent of the integral (if $T=0$)
or finite difference (if $T \ne 0$) equation (\ref{BS2}); this can
be interpreted as a pairing instability (see also
Ref.~[\onlinecite{VG_SDS_KL}]). To unequivocally establish the
existence of a pairing instability for given values of the bare
interactions,
we look for an infrared divergence in Eq.~(\ref{BCSeq}). For this,
we assume that the eigenvector is a weakly dependent function of
the frequency. With this Ansatz, we get
$$
\left[ 1 + (-1)^{l} \nu \kappa U(2p_F)  E_F^{2\sigma -2/3}
\int\limits_T^{E_F} {d \varepsilon \over |\varepsilon|^{2\sigma
+1/3} }\right] \Phi_l = 0,
$$
where $\kappa = (2/3)\Gamma_0^2 B(1/3,2\sigma-1/3)  \left( E_F /
\omega_0\right)^{1/3}$ is just a dimensionless constant, $\nu =
m/\pi$, and we assumed that the Fermi energy serves as a
high-energy cut-off. Clearly a non-trivial solution always exists
if $l$ is an odd number and $\sigma \ge 1/3$. The critical value
$\sigma =1/3$ gives a logarithmic divergence and the transition
temperature $T_* \sim E_F \exp\left\{ - 1/[\nu \kappa U(2p_F)]
\right\}$, while for larger values we get a power-law singularity
and $T_* \sim E_F [\nu \kappa U(2p_F)]^{1/(2\sigma -2/3)}$.

We note that the transition temperature depends on the orbital
angular momentum very weakly (apart from the odd-even dependence);
an explicit dependence on the value of $l$ appears only in the
subleading orders and enters via the overall cut-off constant.
Due to this weak dependence of the transition temperature on $l$,
the actual pairing symmetry is non-universal and depends on the
structure of the effective interaction far from $2 p_F$ and/or
details of the fermion dispersion. What our argument above
explicitly shows is that in the theory with $\sigma \ge 1/3$, the
system is unstable to the pairing at some odd orbital angular
momentum (or momenta). Thus the paired state is a $Z_2$ spin
liquid~\cite{Wen_review}.

If the pairing occurs in a single odd angular momentum channel,
then at some finite temperature of order $T_*$ we expect a
crossover from the spinon-Fermi surface phase to a pseudogap-like
phase as the temperature decreases. On the other hand, the
susceptibility and specific heat coefficients seem finite at low
temperatures below the ``kinks'' at a few Kelvin. There may be two
possible ways to explain this: a)~The low temperature phase is not
a homogeneous phase. Various extrinsic effects (such as tiny
amount of disorder, structural inhomogeneity, and warping of the
spinon Fermi surface) can lead to the suppression of the local
transition temperature and the local pairing amplitude in certain
regions of real space. The resulting state would consist of the
gapless spinon puddles (where the pairing amplitude vanishes) in
the sea of a $Z_2$ spin liquid. This may be consistent with the
observation that two different NMR relaxation times have been
observed [\onlinecite{Shimizu_etal,Kurosaki_etal}]. The finite
susceptibility and specific heat coefficient at low temperatures
(below the ``kink'' temperature) may be due to the gapless
regions.
%
b)~Another possibility is that the pairing occurs in a large
orbital angular momentum channel or the pairing amplitude is given
by a superposition of contributions from multiple angular momentum
channels. Then a large number of nodal excitations may exist and
practically it may look like there is an extended gapless region
in momentum space. Which scenario is realized in real materials
may depend on non-universal physics and it requires more detailed
analysis of the band dispersion and sub-dominant interactions in
the Cooper channel. This is beyond the scope of this paper, but it
may be an excellent topic of future study.


In conclusion, we demonstrated that there exist
Kohn-Luttinger-like pairing instabilities in the $U(1)$ spin
liquid with a spinon Fermi surface. This pairing occurs in
odd-angular-momentum/spin-triplet channels. As explained in the
introduction, this is consistent with several experimental
findings. On the other hand, the previous proposal of the
``Amperian pairing''~\cite{SSLee} (while it explains various
aspects of the same experiments) suggests a spin-singlet pairing
and a spatially modulated pairing amplitude with a well defined
wavevector. This corresponds to an incommensurate version of the
valence bond solid phase and would induce lattice distortion. Thus
the difference between two proposals may be tested by X-ray
scattering.

Finally we emphasize that the microscopic scenario presented
here may be relevant not only to the gauge-fermion system of
spinons but also to other fermionic systems with singular
interactions, which include electrons near an Ising ferromagnetic
instability or near a nematic ordering quantum critical
point~\cite{FMQCP} and the vortex metal phase proposed in
Refs.~[\onlinecite{vm1,vm2,vm3}].


The authors are grateful to  A.~Chubukov, O.~Motrunich, and
G.~Refael for discussions about this work. This research was
supported in part by the NSF under Grant No.~PHY05-51164. It was
also supported by the NSERC, CRC, CIFAR, and KRF-2005-070-C00044
(YBK). The authors acknowledge the KITP for hospitality,  where
most of this work was performed.

\end{document}